\documentclass[twocolumn,prl,showpacs,superscriptaddress,notitlepage,amssymb,preprintnumbers]{revtex4-1}

\usepackage{amsmath,amssymb,amsthm}
\usepackage{braket}
\usepackage{empheq}
\usepackage{enumitem}

\usepackage[pdfstartview=FitH]{hyperref}
\hypersetup{
    colorlinks=true,       
    linkcolor=cyan,          
    citecolor=magenta,        
    filecolor=magenta,      
    urlcolor=cyan,           
    runcolor=cyan
}

\newcommand{\bes} {\begin{subequations}}
\newcommand{\ees} {\end{subequations}}
\newcommand{\ba}{\begin{eqnarray}}
\newcommand{\ea}{\end{eqnarray}}

\newcommand{\bi}{\overline{i}}
\newcommand{\mrp}{\mathrm{p}}
\newcommand\norm[1]{\left\lVert#1\right\rVert}
\newcommand\Tr{\mathrm{Tr}}
\newcommand{\ketbra}[1]{|{#1}\rangle\langle#1|}

\newtheorem{theorem}{Theorem}
\newtheorem*{theorem*}{Theorem}

\newtheorem{condition}{Condition}

\def\a{\alpha}

\begin{document}
\title{Error suppression for Hamiltonian-based quantum computation using subsystem codes}

\author{Milad Marvian}
\affiliation{Department of Electrical Engineering, University of Southern California, Los Angeles, California 90089, USA}
\affiliation{Center for Quantum Information Science \&
Technology, University of Southern California, Los Angeles, California 90089, USA}

\author{Daniel A. Lidar}
\affiliation{Department of Electrical Engineering, University of Southern California, Los Angeles, California 90089, USA}
\affiliation{Department of Physics and Astronomy, University of Southern California, Los Angeles, California 90089, USA}
\affiliation{Center for Quantum Information Science \&
Technology, University of Southern California, Los Angeles, California 90089, USA}
\affiliation{Department of Chemistry, University of Southern California, Los Angeles, California 90089, USA}

\begin{abstract}
We present general conditions for quantum error suppression for Hamiltonian-based quantum computation using subsystem codes. This involves encoding the Hamiltonian performing the computation using an error detecting subsystem code and the addition of a penalty term that commutes with the encoded Hamiltonian. The scheme is general and includes the stabilizer formalism of both subspace and subsystem codes as special cases.
We derive performance bounds and show that complete error suppression results in the large penalty limit. To illustrate the power of subsystem-based error suppression, 
we introduce fully $2$-local constructions for protection against local errors of the swap gate of adiabatic gate teleportation and the Ising chain in a transverse field. 
\end{abstract}
\maketitle


A general strategy for protecting quantum information is to encode this information into a larger system in such a way that the effect of the bath is eliminated, suppressed, or corrected \cite{Lidar-Brun:book}. A promising approach for quantum error suppression in 
Hamiltonian quantum computation \cite{aharonov_adiabatic_2007,Cirac:2012pi,Childs:2013kx} was proposed in Ref.~\cite{jordan2006error}. 
In this scheme one chooses a stabilizer quantum error detection code \cite{Gottesman:1996fk}, encodes the Hamiltonian by replacing each of its Pauli operators 
by the corresponding encoded Pauli operator of the chosen code, and adds penalty terms (elements of the code's stabilizer) that suppress the errors the code is designed to detect. This results in the suppression of excitations out of the ground subspace. By indefinitely increasing the energy scale of the penalty terms this suppression can be made arbitrarily strong \cite{Bookatz:2014uq}.

By construction, this encoding necessitates greater than two-body interactions,
which can make its implementation challenging. An important open question is whether there exist quantum error suppression schemes that involve only two-body interactions. However, even for the special case of quantum memory, invoking penalty terms but no encoding, two-body commuting Hamiltonians cannot in general provide suppression \cite{Marvian:2014nr}. This no-go result left open the possibility that non-commuting two-local Hamiltonians might nevertheless suffice for quantum error suppression. Examples based on (generalized) Bacon-Shor codes \cite{Bacon:05} were recently given  in Ref.~\cite{Jiang:2015kx} to show that this is the case for penalty terms and encoded single-qubit operations, and for some encoded two-qubit interactions, but without general conditions or performance bounds. 

Here we show how general subsystem codes can be used for quantum error suppression. Using an exact, non-perturbative approach, we find conditions that penalty Hamiltonians should satisfy to guarantee 
complete error suppression 
in the infinite energy penalty limit. We derive performance bounds for finite energy penalties. Our formulation accounts for stabilizer subspace and subsystem codes as special cases, including the examples of Refs.~\cite{jordan2006error,Bookatz:2014uq,Jiang:2015kx}. We provide several examples where our approach results in encoded Hamiltonians and penalty terms that involve purely two-body interactions \cite{DD-errsup-comment}.
These examples include the swap gate used in adiabatic gate teleportation \cite{Bacon:2009pr}, and the Ising chain in a transverse field frequently encountered in adiabatic quantum computation and quantum annealing. 

\textit{Setting}.---%
We wish to protect a quantum computation performed by a system with Hamiltonian $H_S(t)$ against the system-bath interaction $V=\sum_j {E_j \otimes B_j}$, to a bath with Hamiltonian $H_B$. 
We construct the encoded system Hamiltonian, $\overline{H}_S(t)$, by replacing every operator in $H_S(t)$ by the corresponding logical operators of a subsystem code \cite{Kribs:2005:180501,Kribs:05,nielsen2007algebraic}. The strategy for protecting the computation performed by $\overline{H}_S(t)$ is to add a penalty Hamiltonian $E_{\mrp} H_{\mrp}$, chosen so that $[\overline{H}_S(t),H_{\mrp}]=0$ in order to prevent interference with the computation~\cite{jordan2006error}. 
As the energy penalty $E_{\mrp}$ is increased, errors should become more suppressed. 

\textit{Results in the infinite penalty limit}.---%
We now state our main results, in the form of two related theorems that give sufficient conditions for complete error suppression in the large $E_{\mrp}$ limit. These results incorporate both those for general stabilizer penalty Hamiltonians introduced in \cite{jordan2006error,Bookatz:2014uq} and the subsystem penalty Hamiltonian examples introduced in \cite{Jiang:2015kx}. They are also related to a dynamical decoupling approach for protecting adiabatic quantum computation \cite{PhysRevLett.100.160506} via a formal equivalence found in Ref.~\cite{Young:13}.

Let $U_0$, $U_{\mrp}$, $U_V$, and $U_W$ be the unitary evolutions generated by $H_0 = \overline{H}_S+H_B$, $E_{\mrp} H_{\mrp}$, $H_V= H_0+E_{\mrp} H_{\mrp} + V$, and $H_W = H_0+E_{\mrp} H_{\mrp} + W$, respectively. 
As will become clear later, $W$ will play the role of the suppressed version of $V$. We assume that $\norm{V},\norm{W}<\infty$, where $\norm{\cdot}$ denotes any unitarily invariant norm \cite{Bhatia:book}.
Let $P$ be an arbitrary projection operator and let $H_{\mrp}=\sum_{a}{\lambda_a \Pi_a}$ be the eigendecomposition of the penalty term.
\begin{theorem} 
\label{thm1}
Set $W=c I$ ($c\in\mathbb{R}$, $I$ is the identity operator) and assume that
\bes
\begin{align}
\label{eq:PH0}
[\overline{H}_S,P] =[\overline{H}_S,H_{\mrp}] &=0 \ , \\
\sum_a \Pi_a V \Pi_a P&=c  P  \ .
\label{eq:thm1cond}
\end{align}
\ees
Then 
\ba
\lim_{E_{\mrp} \rightarrow \infty} \norm{ U_V(T) P -U_W(T) P} = 0 \ ,
\label{eq:thm1}
\ea
where $U_W(T) = e^{-icT} U_0(T)U_{\mrp}(T)$.
\end{theorem}
Theorem~\ref{thm1} states that in the infinite penalty limit and over the support of $P$, the evolution generated by the total system-bath Hamiltonian $H_V$ is indistinguishable (up to a global phase) from the decoupled evolution generated by $H_0+E_{\mrp} H_{\mrp}$. The conditions in Eq.~\eqref{eq:PH0} ensure compatibility of the subspace defined by $P$ and of the type of penalty Hamiltonian $H_{\mrp}$ with the given encoded Hamiltonian $\overline{H}_S$. The condition in Eq.~\eqref{eq:thm1cond} ensures the absence of a term that cannot be removed by the penalty [see the Supplementary Material (SM)].
\begin{theorem} \label{thm:generalsubsys}
Set $W=\sum_{a \in \mathcal{I}}{\Pi_a V \Pi_a}$, where $\mathcal{I}$ is some index set. Assume that in addition to Eq.~\eqref{eq:PH0} also $P=\sum_{a\in\mathcal{I}}{\Pi_a}$. Then Eq.~\eqref{eq:thm1} holds again, with
%
\ba
U_W(T)=\mathcal{T}\exp\int_0^T (H_0(t)+E_{\mrp} H_{\mrp} + W)dt 
\label{eq:U_W(T)}
\ea 
\end{theorem}
($\mathcal{T}$ denotes time-ordering). Theorem~\ref{thm:generalsubsys} is similar to Theorem~\ref{thm1}, except that it allows for a more general target evolution operator $U_W(T)$.
As discussed below, Theorem~\ref{thm1} is suitable for stabilizer subsystem codes, while Theorem~\ref{thm:generalsubsys} is suitable for general subsystem codes. 

\textit{Proof sketch}.---%
Both Theorems~\ref{thm1} and \ref{thm:generalsubsys} establish the desired decoupling result, and show that in principle it is possible to completely protect Hamiltonian quantum computation against coupling to the bath. To prove them we define
\ba
K(t)=\int_0^t {U^\dagger_{\mrp}(\tau)(V -W)U_{\mrp}(\tau) d\tau P} \ ,
\label{eq:K(t)}
\ea
and derive the following bounds in the SM:
\bes
\begin{align}
\label{eq:UV-UW}
&\norm{U_V(T) P- U_W(T) P} \leq \norm{K(T)}  \\
&\,\,\,\, +T\sup_{t}\norm{[K(t),H_0(t)]} +T(\norm{V}+\norm{W})\sup_{t}{\norm{K(t)}} \notag \\
%
&\norm{K(t)} \leq   \frac{2}{E_{\mrp}} \sum_{a\neq a'} \frac{\norm{V-W}}{|\lambda_{a}-\lambda_{a'}|}  \ .
\label{eq:normK}
\end{align}
\ees
Theorems~\ref{thm1} and \ref{thm:generalsubsys} follow in the large $E_{\mrp}$ limit, since in this limit $\norm{K(T)}\to 0$, and $\norm{[K(t),H_0(t)]}\leq 2\|K(t)\|\|H_0\|$. An error bound for finite $E_{\mrp}$ follows directly from Eq.~\eqref{eq:normK} (for related results see Refs.~\cite{Bookatz:2014uq,Marvian:2015sf}). While a tighter bound may not be possible without introducing additional assumptions, we note that for a Markovian bath in a thermal state, it is possible to show that the excitation rate out of the code space is exponentially suppressed as a function of $E_{\mrp}$, and $E_{\mrp}$ need only grow logarithmically in the system size to achieve a constant excitation rate, assuming the gap of $H_\mrp$ is constant \cite{jordan2006error,marvian2016error}.

\textit{Subsystem codes}.---%
Before demonstrating the implications of Theorems~\ref{thm1} and \ref{thm:generalsubsys} we first briefly review subsystem codes. 
Assume that the system's Hilbert space can be decomposed as $\mathcal{H}_S=C \oplus C^\perp$, where $C = {A} \otimes {B}$. 
The channel (completely positive map) $\mathcal{E}  =\{E_j\}$ is 
\emph{detectable} on the ``information subsystem" $B$ if (see the SM for a proof):
\ba 
\label{Knill-Laflamme}
\forall E_j \, \exists G_{j}: P_C E_j P_C = P_C G_{j} \otimes I_B P_C\ ,
\ea
where $I_B$ is the identity on $B$ and $P_C$ denotes the projector onto $C$. Here $A$ plays the role of a ``gauge subsystem"; the $G_{j}$ operators are arbitrary and do not affect the information stored in subsystem $B$. 

Stabilizer subsystem codes \cite{poulin_stabilizer_2005} are of particular interest. Intuitively, one can think of such codes as subspace stabilizer codes  \cite{Gottesman:1996fk} where some logical qubits and the corresponding logical operators are not used. 
A stabilizer code can be defined as the subspace stabilized by an Abelian group $\mathcal{S}=\braket{S_1,...,S_s}$ of Pauli operators, with $-I \notin \mathcal{S}$, where $\{S_i\}_{i=1}^s$ are the group generators. The projector onto the codespace is $P_{C}=\prod_{i=1}^s \frac{I+ S_i}{2}$. To induce a subsystem structure we define logical operators $\mathcal{L}$ and gauge operators $\mathcal{A}$ as Pauli operators that leave the codespace invariant, and also demand that the three sets $\mathcal{S}$, $\mathcal{L}$, and $\mathcal{A}$ mutually commute. The generators of $\mathcal{L}$ and $\mathcal{A}$ can be organized in canonical conjugate pairs: the set of bare logical operators $\mathcal{L}=\{\overline{Z}_1,\overline{X}_1,...,\overline{Z}_k,\overline{X}_k\}$ that preserve the code space and
act trivially on the gauge qubits \cite{PhysRevA.83.012320}, and the set of gauge operators $\mathcal{A}=\{ Z'_1,X'_1,...,Z'_r,X'_r\}$, where for $A,B \in\{\overline{X},\overline{Z}\}$ or $A,B \in\{X',Z'\}$ we have $[A_i,B_j]=0$ if $i\neq j$, and $\{\overline{X}_i,\overline{Z}_i\}=0$. The gauge group is defined as $\mathcal{G}=\braket{S_1,...,S_s,Z'_1,X'_1,...,Z'_r,X'_r}$, and is non-Abelian. 
A Pauli error $E_j$ is detectable iff it anti-commutes with at least one of the stabilizer generators \cite{poulin_stabilizer_2005}, or equivalently iff $P_C E_j P_C=0$ [since 
$(I+ S_i)(I- S_i)=0$ $\forall i$].

%

\textit{Protection using stabilizer codes}.---%
To satisfy the condition $[\overline{H}_S,H_{\mrp}] =0$ in Eq.~\eqref{eq:PH0} we may choose $H_{\mrp}$ as a linear combination of elements of the gauge group $\mathcal{G}$ (not necessarily the generators) \cite{Jiang:2015kx,AQC-DD-comment-gauge},
\ba 
H_{\mrp}=\sum_{i} \alpha_i g_i\ , \quad g_i \in \mathcal{G}\ , \quad |\alpha_i| \leq 1 \ , \forall i\ . 
\label{def:Hp}
\ea
To satisfy the condition $[\overline{H}_S,P] =0$ we may choose
$P=\sum_{a \in \mathcal{I}}\Pi_a$. Equation~\eqref{eq:thm1cond} then becomes 
$\Pi_a V \Pi_a=c \Pi_a$ $\forall a \in \mathcal{I}$, 
a condition that is already satisfied with $c=0$ for a stabilizer error detecting code (for which $P_C V P_C=0$) if the support of $P$ is in the codespace (i.e., $PP_C=P_C P=P$). 
This is true, in particular, if $\mathcal{I}$ contains just the ground subspace of $H_{\mathrm p}$.
We may thus state the following corollary of Theorem~\ref{thm1}: \emph{For $H_{\mrp}$ chosen as in Eq.~\eqref{def:Hp}, the joint system-bath evolution completely decouples in the large penalty limit for initial states in the ground subspace of $H_{\mrp}$, with this subspace itself being a subspace of the codespace.} 

Note that the difference between the subspace and subsystem case manifests itself in the appearance of $U_{\mrp}(T)$ in Eq.~\eqref{eq:thm1}. If the penalty Hamiltonian consisted of only stabilizer terms [i.e., $g_i\in\mathcal{S}$ $\forall i$ in Eq.~\eqref{def:Hp}], the penalty Hamiltonian would at most change the overall phase of states in the codespace. But here, as the elements of penalty Hamiltonian can be any element of the gauge group, $U_{\mrp}$ can have a nontrivial effect on states in $C$.
Nevertheless, as the gauge operators commute with the logical operators of the code, this unitary does not change the result of a measurement of the logical subsystem.  In the SM we provide a formal argument using a distance measure to quantify state distinguishability using generalized measurements restricted to the logical subsystem.

\textit{Protection using general  subsystem codes}.---%
Choose a code $C$ with projector $P_C$ such that the error-detection condition~\eqref{Knill-Laflamme} is satisfied for all the error operators $\{E_j\}$ in $V=\sum_j {E_j \otimes B_j}$. Assume that the penalty is chosen so that $[\overline{H}_S,H_{\mrp}] =0$ in Eq.~\eqref{eq:PH0} holds, and set $P=P_C$ in Theorem~\ref{thm:generalsubsys} (thus also the condition $[\overline{H}_S,P] =0$ holds). Then $\Pi_a V \Pi_a=\sum_j (\Pi_a G_{j} \otimes I_B \Pi_a) \otimes B_j$ $\forall a\in\mathcal{I}$, so that $W=\sum_{a \in \mathcal{I}}\sum_j (\Pi_a G_{j} \otimes I_B \Pi_a) \otimes B_j$, with trivial action ($I_B$) on the information subsystem $B$. The unitary $U_W(T)$ [Eq.~\eqref{eq:U_W(T)}] appearing in Theorem~\ref{thm:generalsubsys} thus has a non-trivial effect on $B$ only via  the $H_0(t)$ term, as desired.

\textit{Block encoding}.---%
A useful simplification results when the logical qubits can be partitioned into $n$ separate blocks. In this case the total penalty Hamiltonian becomes
$H_{\mrp}=\sum_{\bi=1}^{n}{h_{\mrp}^{\bi}}$,
where $h_{\mrp}^{\bi}= \sum_j \alpha^{\bi}_j g^{\bi}_j$ denotes the penalty Hamiltonian on logical qubit $\bi$, with $g^{\bi}_j\in \mathcal{G}$, and $[h_{\mrp}^{\bi},h_{\mrp}^{\overline{j}}]=0$ for $i\neq j$. The code space projector becomes $P_C=\otimes_{i=1}^n p^{\overline{i}}$, where $p^{\overline{i}}$ is the projector onto the code space of the $i$th logical qubit. 
We may also partition the system-bath interaction according to the logical qubits it acts on: $V=\sum_{\bi=1}^{n}{v^{\bi}}$ (note that we do not assume that $[v^{\bi},v^{\overline{j}}]=0$).
%
Clearly, $K(t)$ can also be expressed as a sum over blocks, as can inequality~\eqref{eq:UV-UW}. Using the eigendecomposition $h^{\bi}_{\mrp}=\sum_a {e^{\bi}_a \pi^{\bi}_a}$, 
condition~\eqref{eq:thm1cond} can then be replaced by 
\ba
\pi^{\bi}_a v^{\bi} \pi^{\bi}_a p^{\bi}=c^{\bi} p^{\bi} \quad \forall a,\bi\ .
\label{eq:pi} 
\ea
%
Using the block encoding structure, in the SM we tighten the error bound resulting from Eq.~\eqref{eq:normK}.
We show, in particular, that the bound is extensive in the system size and depends only on the bath degrees of freedom that couple locally to the system, so that the bound is not extensive in the bath size.

\textit{A simplified sufficient condition}.---%
To check whether Theorem~\ref{thm1} applies one can simply find the eigendecomposition of $h_{\mrp}^{\bi}$ and check if Eq.~\eqref{eq:pi} holds for a given system-bath interaction and choice of code space. Instead, we next identify conditions that are less general but are easier to check. We assume that the interaction Hamiltonian has the $1$-local form $V=\sum_{\bi} v^{\bi}$, where $v^{\bi}=\sum_j{\sigma^{\bi}_j \otimes B^{\bi}_j}$ and $\sigma^{\bi}_j $ is an arbitrary non-identity Pauli operator acting on qubit $j$ in block $\bi$. From now on we drop the block superscript for notational simplicity. Furthermore, we choose a penalty term that satisfies $[h_{\mrp},p]=0$ given a code block projector $p$, which implies $[\pi_a,p]=0$ $\forall a$. 

A sufficient condition for Eq.~\eqref{eq:pi}, and hence for Theorem~\ref{thm1}, is then the following: 
\begin{condition}
\label{cond1}
$h_{\mrp}p$ and $\sigma_j h_{\mrp} \sigma_j p$ do not share an eigenvalue for any $\sigma_j$ in the support of $p$.
\end{condition}
To see that this is a sufficient condition, we note that $\pi_a p$ and $\sigma_j \pi_a \sigma_j p$ are both projectors, corresponding to the same eigenvalue $e_a$ of $h_{\mrp}$ and $\sigma_j h_{\mrp} \sigma_j$.
If both projectors are nonzero then there exists at least one (nonzero) eigenvector for each of $h_{\mrp} p$ and $\sigma_j h_{\mrp} \sigma_j p$ with eigenvalue $e_a$, in contradiction to our condition. So, the stated condition guarantees that for any eigenvalue $e_a$ we have either $\pi_a p=0$ or $\sigma_j \pi_a \sigma_j p=0$. Thus, $\forall a$: $0=(\sigma_j \pi_a \sigma_j p)(\pi_a p)= \sigma_j \pi_a \sigma_j p p \pi_a= \sigma_j \pi_a \sigma_j p \pi_a = \sigma_j (\pi_a \sigma_j \pi_a p)$, so that, $\forall a$: $\pi_a \sigma_j \pi_a  p=0$, which implies Eq.~\eqref{eq:pi} {(with $c^{\bi}=0$ $\forall i$)}. 
We now consider a number of interesting cases, and show that Condition~\ref{cond1} holds, thus guaranteeing error suppression via Theorem~\ref{thm1}.

\textit{Stabilizer penalty Hamiltonians}.--- As in Ref.~\cite{jordan2006error}, let
\ba
h_{\mrp}=\sum_i{\alpha_i S_i}
\ea
with $S_i\in \mathcal{S}$, $\alpha_i\neq 0$ and $p=p_c$. Clearly $[h_{\mrp},p]=0$.
Let us define $a_{ij}=0$ or $1$ if $[S_i,\sigma_j]=0$ or $\{S_i,\sigma_j\}=0$, respectively. In the support of $p$ (i.e., in the code space) $h_{\mrp} p =(\sum{\alpha_i}) p$, so the eigenvalue of $h_{\mrp}$ there equals $\sum_i{\alpha_i}$, while the eigenvalue of $\sigma_j h_{\mrp} \sigma_j p$ there equals $\sum_i{\alpha_i (-1)^{a_{ij}}}$. Condition~\ref{cond1} thus requires $\forall j$: $\sum_i{\alpha_i} \neq \sum_i{\alpha_i (-1)^{a_{ij}}}$. When all $\alpha_i$ 
have the same sign
this becomes 
the familiar error detection condition, that every $\sigma_j$ anticommutes with at least one of the terms in the sum of stabilizers.

The penalty Hamiltonian considered in Ref.~\cite{Bookatz:2014uq} corresponds to $h_{\mrp}= I-p$, so that $[h_{\mrp},p]=0$ holds. Condition~\ref{cond1} is also satisfied in this case since since $h_{\mrp}p=0$, while $\sigma_j h_{\mrp} \sigma_j p = p-\sigma_j p \sigma_j p = p$ (where we used the error detection condition $p \sigma_j p=0$), so in the support of $p$ the eigenvalues are, respectively, $0$ and $1$.

\textit{Gauge group penalty Hamiltonians}.---%
A family of generalized Bacon-Shor codes can be identified with a binary matrix $A$, which fully characterizes all the code properties \cite{PhysRevA.83.012320}. E.g., each nonzero element of $A$ corresponding to a qubit on a planar grid, and two ones in a row (column) of the matrix correspond to an $XX$ ($ZZ$) generator acting on the corresponding qubits (see the SM for more details). As pointed out in Ref.~\cite{Jiang:2015kx}, because of the locality of the generators of these codes, they are promising candidates for use in error suppression schemes. We present several examples for suppressing local errors that originate from this construction.

(i) The $[[4,1,2]]$ code was proposed in Ref.~\cite{Jiang:2015kx} to overcome the aforementioned no-go theorem for error suppression using $2$-local commuting Hamiltonians \cite{Marvian:2014nr}. Each qubit is encoded into four qubits using this code (block encoding), so the entire code corresponds to a block diagonal $A$ matrix, with $2\times 2$ blocks of all ones. 
The stabilizer, gauge and bare logical generators are:
\bes
\begin{align}
\mathcal{S}&=\braket{S_1=X^{\otimes 4}, S_2=Z^{\otimes 4}} \\
\mathcal{A}&=\{X'=X_1 X_2,Z'=Z_1 Z_3\} \\
\mathcal{L}&=\{\overline{X}=X_1 X_3,\overline{Z}=Z_1 Z_2\} \ . 
\end{align}
\ees
Thus $\mathcal{G}=\braket{S_1,S_2,X',Z'}=\braket{S_1X',S_2Z',X',Z'}=\braket{X_3 X_4,Z_2 Z_4,X_1 X_2, Z_1 Z_3} \equiv \braket{\{g_i\}_{i=1}^4}$, i.e., the generators are $2$-local. The  penalty Hamiltonian is 
$h_{\mrp}=E_{\mrp} \sum_{i=1}^4  g_i$ and again, clearly $[h_{\mrp},p]=0$. One may check that the eigenvalues of $h_{\mrp}p$ and $\sigma_j h_{\mrp} \sigma_j p$ are $0,\pm 2E_{\mrp}$ and $\pm 2\sqrt{2}E_{\mrp} $, respectively (see the SM).
Thus Condition~\ref{cond1} is satisfied. While the penalty Hamiltonian is $2$-local, unfortunately the encoding of a $2$-local interaction (which is necessary for universal quantum computation), still requires $4$-local interactions. 

(ii) We show how to encode and protect the adiabatic swap gate introduced in \cite{Bacon:2009pr} using purely $2$-local interactions. This Hamiltonian is one of the key building blocks of a proposal for universal quantum computation using adiabatic gate teleportation. The Hamiltonian is:
$H(s)=(1-s)(X_b X_c+ Z_b Z_c) + s (X_a X_b+ Z_a Z_b)$.
By slowly increasing $s$ from $0$ to $1$ any state initially prepared on qubit $a$ transfers onto qubit $c$. 
To encode and protect this Hamiltonian, we use the following $[[8,3,2]]$ subsystem code:
\bes
\begin{align}
\mathcal{S}&=\braket{S_1=X^{\otimes 8}, S_2=Z^{\otimes 8}} \\
\mathcal{L}&=\{ \overline{X}_1=X_1 X_8, \overline{X}_2=X_1 X_2 X_3 X_8, \overline{X}_3=X_4 X_5, \notag \\
& \qquad\overline{Z}_1=Z_1 Z_2, \overline{Z}_2=Z_3 Z_4 Z_5 Z_6, \overline{Z}_3=Z_5 Z_6 \} \\
\mathcal{G}&=\braket{X_1 X_2, \! X_3 X_4, \! X_5 X_6, \! X_7 X_8, \! Z_2 Z_3, \! Z_4 Z_5, \! Z_6 Z_7, \! Z_8 Z_1} 
\notag
\end{align}
\ees
%
The penalty Hamiltonian is the sum of all the gauge group generators $g_i \in \mathcal{G}$, which is manifestly $2$-local. One can check that Condition~\ref{cond1} is satisfied for this Hamiltonian (see the SM), and so we obtain the desired protection.
The encoded Hamiltonian becomes:
\begin{align}
\!\!\!\!\overline{H}(s) &= (1-s)(\overline{X}_2 \overline{X}_3+ \overline{Z}_2 \overline{Z}_3) + s (\overline{X}_1 \overline{X}_2+ \overline{Z}_1 \overline{Z}_2) \notag \\
&=(1-s)(X_6 X_7+ Z_3 Z_4) + s (X_2 X_3+ Z_7 Z_8) ,
\label{eq:2loc-AGT}
\end{align}
where in the second line we used the fact that $\overline{X}_2 \overline{X}_3=S_1 X_6 X_7$ and $ \overline{Z}_1 \overline{Z}_2=S_2 Z_7 Z_8$ are equivalent logical operators. Thus, the encoded Hamiltonian remains $2$-local.



(iii) Our next example, an open Ising chain in a transverse field, does not involve block encoding:
\ba
H_S(s)=(1-s) \sum_{i=1}^N{X_i}+ s \sum_{i=1}^{N-1}J_i {Z_i Z_{i+1}}\ .
\ea
This Hamiltonian appears frequently in adiabatic quantum optimization. The goal is again to provide encoding and error suppression using only $2$-local Hamiltonians. 

Using an $A$-matrix derived $[[2N+2,N,2]]$ code (see the SM for details), we obtain:
\begin{align}
\label{eq:HpforChain}
H_{\mathrm p}&= -\sum_{i=1}^{N+1} {X_{2i-1} X_{2i}} + \sum_{i=1}^{N} {Z_{2i} Z_{2i+1}+Z_1 Z_{2N+2}}  \nonumber\\
\overline{H}_S(s)
&=(1-s) \sum_{i=1}^{N} {X_{2i} X_{2i+1}}+ s \sum_{i=1}^{N-1} J_i {Z_{2i+1} Z_{2i+2}}\ .
\end{align}
We have verified numerically that the ground subspace of $H_{\mathrm p}$ is a subspace of the codespace, which as we showed above is sufficient for error suppression in the stabilizer case. 
We also find numerically that the minimum gap of $H_{\mathrm p}$ decreases as $1/(N+1)$ (see the SM), so that $E_\mathrm{p}$ should grow with $N$ to maintain the protection obtained in this case as the system size increases, since this gap separates the logical ground subspace from the undecodable excited states. While in general this is undesirable, it is compatible with examples where $H_S$ (and hence also $\overline{H}_S$) exhibits more rapidly closing gaps for certain choices of the couplings $\{J_i\}$ (e.g., an exponentially small gap \cite{Reichardt:2004}). 

\textit{Non-additive codes}.---%
Theorems~\ref{thm1} and \ref{thm:generalsubsys} allow us to go beyond the framework of Ref.~\cite{Bookatz:2014uq} and examples of Ref.~\cite{Jiang:2015kx}, and employ non-additive codes (also known as non-stabilizer codes) to encode and protect evolutions \footnote{Ref.~\cite{Bookatz:2014uq} proved a less general version of Theorem~\ref{thm1}, where Eq.~\eqref{eq:thm1cond} is replaced by $P_C V P_C=0$; this excludes non-additive codes. Ref.~\cite{Jiang:2015kx} used certain stabilizer subsystem codes but did not consider non-additive codes.}. 
Non-additive codes can achieve higher rates (ratio of the number of encoded to physical qubits) than stabilizer codes~\cite{rains1997nonadditive,smolin2007simple,cross_codeword_2009,shin_codeword-stabilized_2012}. For example, using $5$ physical qubits to detect any single-qubit error 
stabilizer codes can encode at most $2$ qubits, but using a non-additive code one can encode up to $\log_2 6$ qubits~\cite{rains1997nonadditive}.
The encoding procedure is straightforward. Choosing a subspace code $C$, one can expand the system Hamiltonian in a basis $\{\ket{i}\}$ and then replace each basis vector in the expansion with the corresponding code state $\{\ket{\bar{i}}\}$. One possible choice of a penalty Hamiltonian is $E_{\mrp} H_{\mrp}$, where $H_{\mrp}=-P_C$ and $P_C=\sum_{i \in {C}}{\ketbra{\bar{i}}}$. Theorem~\ref{thm1} guarantees that with this choice, starting from an initial state in the codespace, leakage out of the codespace is suppressed in the large $E_{\mrp}$ limit, and the desired system Hamiltonian is implemented in the codespace with a higher rate than what could be achieved using stabilizer codes. Moreover, Theorem~\ref{thm:generalsubsys} allows using non-additive subsystem codes such as the codes introduced in Ref.~\cite{shin_codeword-stabilized_2012}.

\textit{Conclusions}.---%
We have presented conditions guaranteeing error suppression for Hamiltonian quantum computation using general subsystem error detecting codes, along with conditions that the corresponding penalty Hamiltonians should satisfy, and performance bounds that improve monotonically with increasing energy penalty. 
Stabilizer subsystem codes are more flexible than stabilizer subspace codes when there are constraints on the spatial locality of the generators of the code \cite{PhysRevA.83.012320}. 
This allowed us to use these codes to present examples of fully $2$-local encoded Hamiltonian quantum information processing with error suppression. This should hopefully pave the way towards a similar result for protected universal Hamiltonian quantum computation.

\textit{Acknowledgments}.---%
We thank Todd Brun and Iman Marvian for useful comments. This work was supported under ARO grant number W911NF-12-1-0523, ARO MURI Grant Nos. W911NF-11-1-0268 and W911NF-15-1-0582, and NSF grant number INSPIRE-1551064.

\bibliography{refs}

\begin{thebibliography}{34}
\bibitem [{DD-()}]{DD-errsup-comment}%
  \BibitemOpen
  \href@noop {} {}\bibinfo {note} {{We note that using dynamical decoupling it
  is possible to enact protected universal encoded adiabatic quantum
  computation using purely two-body interactions in the stabilizer subspace
  formalism \cite{PhysRevLett.100.160506}}}\BibitemShut {NoStop}%
\bibitem [{AQC()}]{AQC-DD-comment-gauge}%
  \BibitemOpen
  \href@noop {} {}\bibinfo {note} {{This penalty Hamiltonian is analogous
  \cite{Young:13} to the Hamiltonian generating the dynamical decoupling pulses
  in the approached introduced in
  Ref.~\cite{PhysRevLett.100.160506}.}}\BibitemShut {Stop}%
\end{thebibliography}

\newpage

\onecolumngrid

\appendix

\section{Proof of Theorem 1 and Theorem 2, and derivation of Eqs. (5a) and (5b)} 
\label{Mainthm}

\begin{proof}

Let $H_V= H_0+E_{\mrp} H_{\mrp} + V$ and $H_W = H_0+E_{\mrp} H_{\mrp} + W$, with corresponding unitary evolutions $U_V$ and $U_W$, respectively. 
We would like to bound $\norm{U_V(T) P- U_W(T) P}$, where $T$ is the final time. Below we suppress the time dependence for notational simplicity unless it is essential. Only $V$, $W$, and $H_{\mrp}$ are time-independent. 

Going to the interaction picture defined by $H_0+E_{\mrp} H_{\mrp}$, we let $\tilde{V}=U_0^{\dagger}U_{\mrp}^{\dagger} V U_{\mrp} U_0$ and $\tilde{W}=U_0^{\dagger}U_{\mrp}^{\dagger} W U_{\mrp} U_0$. We denote the corresponding unitary evolutions by $\tilde{U}_V$ and $\tilde{U}_W$. 

Now, for any unitarily invariant norm \cite{Bhatia:book}, in particular the operator norm:
\bes
\ba
\norm{U_V P- U_W P}&=&\norm{U_0 U_{\mrp} \tilde{U}_V P- U_0 U_{\mrp} \tilde{U}_W P}=\norm{\tilde{U}_V P-\tilde{U}_W P}=\norm{(\tilde{U}_V^{\dagger} \tilde{U}_W-I) P}\\
&=&\norm{\int_0^T \frac{d [\tilde{U}_V^{\dagger}(\tau) \tilde{U}_W(\tau)]}{d \tau} d\tau P}
=\norm{\int_0^T \tilde{U}_V^{\dagger} [\tilde{V}-\tilde{W}] \tilde{U}_W d\tau P}\\
&=&\norm{\int_0^T \tilde{U}_V^{\dagger} U_0^{\dagger}U_{\mrp}^{\dagger} (V-W) U_{\mrp} U_0 \tilde{U}_W d\tau P} \ .
\ea
\ees
We assume that $[P,\tilde{W}] = 0$ and verify this condition below. Then:
\ba
\norm{U_V P- U_W P}=\norm{\int_0^T \tilde{U}_V^{\dagger} U_0^{\dagger}U_{\mrp}^{\dagger} (V-W) U_{\mrp} P  U_0 \tilde{U}_W d\tau P} \ .
\ea
Recall that $K(t)=\int_0^t {U^\dagger_{\mrp}(\tau)(V -W)U_{\mrp}(\tau) d\tau P}
$ [Eq. (4)]. Using integration by parts,\footnote{The version we need here is the following: let $A$, $B$ and $C$ be operators. Then $(ABC)' = aBC+AbC+ABc$ where $a=A'$ etc. By the fundamental theorem of calculus $\int_r^s (ABC)' = ABC|_r^s = \int_r^s (aBC+AbC+ABc)$. We set $A=\tilde{U}_V^{\dagger} U_0^{\dagger}$, $b=\frac{d K}{dt}$, and $C=U_0 \tilde{U}_W$.} we have:
\bes
\ba
U_V P- U_W P &=&\int_0^T \tilde{U}_V^{\dagger} U_0^{\dagger}\frac{d K}{dt}  U_0 \tilde{U}_W dt P  \\
&=&\tilde{U}_V^{\dagger}(T) U_0^{\dagger}(T) K(T) U_0(T)  \tilde{U}_W(T) P - \int_0^T (i\tilde{U}_V^{\dagger} \tilde{V}) U_0^{\dagger} K U_0  \tilde{U}_W dt P    \\
&-& \int_0^T \tilde{U}_V^{\dagger} U_0^{\dagger} K U_0  (-i\tilde{W} \tilde{U}_W) dt P 
-i\int_0^T \tilde{U}_V^{\dagger} U_0^{\dagger} [H_0, K] U_0  \tilde{U}_W dt P \ . 
\ea
\ees
Using the triangle inequality, unitary invariance, and submultiplicativity we thus have the upper bound:
\ba
\norm{U_V(T) P- U_W(T) P} \leq \norm{K(T)}+T(\norm{V}+\norm{W})\sup_{t}{\norm{K(t)}}+T\sup_{t}\norm{[K(t),H_0(t)]} \ ,
\label{eq:UV-UW-again}
\ea
which is Eq. (5a).

Next, let us substitute the eigendecomposition $H_{\mrp}=\sum_{a}{\lambda_a \Pi_a}$ into $K(t)$:
\ba
K(t)=  \int_0^t{ \sum_{a\neq a'}{e^{i (\lambda_{a}-\lambda_{a'}) E_{\mrp} \tau} \Pi_{a} (V-W)  \Pi_{a'}}d\tau P} +\int_0^t{ \sum_{a}{ \Pi_{a} (V-W)  \Pi_{a}}d\tau P} \ .
\label{eq:K(t)2}
\ea

The last term vanishes under the assumptions of Theorems 1 and 2. To see this, note that Theorem 1 corresponds to the case where $W=c I$, so also $\tilde{W} = cI$ and $[P,\tilde{W}] = 0$ is satisfied. Additionally, $\sum_a \Pi_{a} (V-W)  \Pi_{a}P$ becomes $\sum_{a}{ \Pi_{a} V  \Pi_{a}}P=c P$, which is satisfied by assumption in Theorem 1. 

Similarly, Theorem 2 corresponds to the case where $W=\sum_{a \in \mathcal{I}}{\Pi_a V \Pi_a}$ and $P=\sum_{a \in \mathcal{I}}{\Pi_a}$, so again $[P,\tilde{W}] = 0$ is satisfied. In this case too, $\sum_{a}{ \Pi_{a} (V-W)  \Pi_{a}}P=0$.

Carrying out the remaining integral in Eq.~\eqref{eq:K(t)2}, we thus obtain:
\bes
\label{eq:normK2}
\ba
\norm{K(t)} &=&\norm{ \sum_{a\neq a'}{\frac{e^{i (\lambda_{a}-\lambda_{a'}) E_{\mrp} t}-1}{(\lambda_{a}-\lambda_{a'}) E_{\mrp}} \Pi_{a} (V-W)  \Pi_{a'}}P} 
\leq \frac{1}{E_{\mrp}} \sum_{a\neq a'}{\frac{2}{|\lambda_{a}-\lambda_{a'}|} \norm{ \Pi_{a} (V-W)  \Pi_{a'}P}} \\
& \leq  & \frac{2}{E_{\mrp}} \sum_{a\neq a'} \frac{\norm{V-W}}{|\lambda_{a}-\lambda_{a'}|}  
\stackrel{E_{\mrp}\to\infty}{\longrightarrow} 0 \ ,
\ea
\ees
which confirms Eq.~(5b).

\end{proof}

\subsection{Bounding the error for finite penalties assuming block encoding}
\label{app:norm-blocks}

In this section we show that the bound resulting from Eq. (5b) for the finite penalty case can be tightened, and in particular does not depend extensively on the bath size via $\norm{V-W}$ if the bath couples locally to the system.

For simplicity we assume a block encoding with one logical qubit per block. The same method can be used when several logical qubits are encoded in each block.  Let us expand $V$ and $W$ as $V=\sum_{\bi=1}^{n}\sum_{j \in \bi } v^{\bi}_j$ and $W=\sum_{\bi=1}^{n}\sum_{j \in \bi } w^{\bi}_j$. The first sum is over the logical qubits, the second is over terms with support on a logical qubit.\footnote{E.g., assuming local terms act on system sites, and encoding each logical qubit into $n_c$ physical qubits, $j$ in the second sum runs from $1$ to $3n_c$ for each $\bi$.} Accordingly we define:
\ba \label{defineKj}
K^{\bi}_j(t)=\int_0^t{U^\dagger_{\mrp}(\tau) (v^{\bi}_j-w^{\bi}_j)  U_{\mrp}(\tau) P d\tau}\ .
\ea
With the penalty Hamiltonian represented as a sum over logical qubits, $H_{\mrp}=\sum_{\bi=1}^{n}{h_{\mrp}^{\bi}}$, we have:
\ba 
K^{\bi}_j(t)=\int_0^t{e^{i h^{\bi}_{\mrp}  \tau} (v^{\bi}_j-w^{\bi}_j)  e^{-ih^{\bi}_{\mrp} \tau} P d\tau}\ .
\ea
Thus, using the general bound~\eqref{eq:UV-UW-again}:
\ba
\norm{U_V(T) P- U_W(T) P} \leq \sum_{\overline{i}}  \sum_{j \in \bi} \norm{K^{\bi}_j(T)}+T(\norm{V}+\norm{W})\sup_{t}{\norm{K^{\bi}_j(t)}}+T\sup_{t}\norm{[K^{\bi}_j(t),H_0]}\ .
\ea
The only part of $H_0(t)= \overline{H}_S(t)+H_B$ that appears in the bound is that which does not commute with $K^{\bi}_j$:
\bes
\begin{align} 
[K^{\bi}_j(t),H_0(t)]&=\int_0^t{e^{i h^{\bi}_{\mrp}  \tau} [v_j-w_j,H_0(t)]  e^{-i h^{\bi}_{\mrp}  \tau}P d\tau}=\int_0^t{e^{i h^{\bi}_{\mrp}  \tau} [v^{\bi}_j-w^{\bi}_j,\overline{h}^{\bi}_{S,j}(t)+h^{\bi}_{B,j}]  e^{-i h^{\bi}_{\mrp}  \tau} P d\tau}\\
&= [K^{\bi}_j(t),\overline{h}^{\bi}_{S,j}(t)+h^{\bi}_{B,j}] \ ,
\end{align}
\ees
where $\overline{H}_S(t) = \sum_{\overline{i}}  \sum_{j \in \bi}\overline{h}^{\bi}_{S,j}(t)$ and $H_B = \sum_{\overline{i}}  \sum_{j \in \bi}h^{\bi}_{B,j}$, with $\overline{h}^{\bi}_{S,j}(t)$ representing the part of the system that has nontrivial support on site $j$ in block $\bi$ of the system and $h^{\bi}_{B,j}$ representing the bath Hamiltonian part that has nontrivial support on the bath part of the interaction Hamiltonian, here corresponding to $v^{\bi}_j-w^{\bi}_j$ . Using the triangle inequality and unitary invariance of the norm we have:
\ba
\!\!\!\!\!\!\!\!\!\!\!\!\norm{U_V(T) P- U_W(T) P} \leq \sum_{\overline{i}}  \sum_{j \in \bi} \|{K^{\bi}_j(T)}\|+T(\norm{V}+\norm{W})\sup_{t}{\|{K^{\bi}_j(t)}\|}+2T\sup_{t} \|{K^{\bi}_j(t)}\|(\|\overline{h}^{\bi}_{S,j}(\tau)\|+\| h^{\bi}_{B,j}\|) \ .
\label{eq:localboundU-V}
\ea
This involves the local bath component $h^{\bi}_{B,j}$, as opposed to depending extensively on $\| H_B\|$.

The only remaining ingredient is: 
\ba
\norm{K^{\bi}_j(t)}=\norm{\int_0^t{e^{+i h^{\bi}_{\mrp}  \tau}(v^{\bi}_j-w^{\bi}_j)  e^{-ih^{\bi}_{\mrp} \tau} p^{\bi} d\tau}}\ ,
\ea
and we repeat the steps leading from Eq.~\eqref{eq:K(t)2} to Eq.~\eqref{eq:normK2}. Namely, using the eigendecomposition $h^{\bi}_{\mrp}=\sum_a {e^{\bi}_a \pi^{\bi}_a}$ and assuming that 
\ba
\pi^{\bi}_a (v^{\bi}_j-w^{\bi}_j) \pi^{\bi}_a p^{\bi}=0 \quad \forall a,\bi,j 
\label{eq:pi2} 
\ea
holds [a generalization of Eq.~(8) allowing for $W\neq 0$] so that the term corresponding to the second integral in Eq.~\eqref{eq:K(t)2} vanishes, the bound becomes:
\ba
\norm{K^{\bi}_j(t)} \leq \sum_{a\neq a'} \norm{\pi^{\bi}_a (v^{\bi}_j-w^{\bi}_j)  \pi^{\bi}_{a'} p^{\bi}}  \frac{|e^{+i E_{\mrp} (e^{\bi}_a-e^{\bi}_{a'})t}-1|}{ |E_{\mrp} (e^{\bi}_a-e^{\bi}_{a'})|}
\leq \frac{2}{E_{\mrp}}\sum_{a\neq a'}   \frac{\|{v^{\bi}_j-w^{\bi}_j}\|}{ |e^{\bi}_a-e^{\bi}_{a'}|} \ .
\label{Bound_K}
\ea
Thus, the bound for the finite penalty case only depends locally on the coupling to the bath, via $\|{v^{\bi}_j-w^{\bi}_j}\|$. As expected, the bound remains extensive in the system size, via the sum over $\bi$ in Eq.~\eqref{eq:localboundU-V}.

\section{Proof of the subsystem error detection condition, Eq.~(6)}

For completeness, we provide a proof of the sufficiency of the error detection condition (see Ref.~\cite{Kribs:2005:180501,Kribs:05,nielsen2007algebraic} for necessary and sufficient conditions and proofs for correctable errors on a subsystem.)
The channel $\mathcal{E}  =\{E_i\}$ is detectable by a code $C$ if there exists a measurement that unambiguously reveals whether or not an error took place after $\mathcal{E}$ acts on a state $\ket{\bar{\psi}_\alpha}\in C$, $\forall \alpha$. For subsystem codes, states in $C$ are allowed to change by a gauge transformation. To show that Eq. (6) is sufficient for error detection we rewrite it as
\begin{equation}
E_i\ket{\bar{\psi}_{\a}}=G_{i} \otimes I_B\ket{\bar{\psi}_{\a}}+\ket{\phi^{\bot}_{{\a},i}}\ \forall i,\alpha\ ,
\label{eq:erEq}
\end{equation}
for some (unnormalized) state $\ket{\phi^{\bot}_{{\a},i}}\in\mathcal{C}^{\bot}$.

The action of the channel is then
\bes
\begin{align} 
\mathcal{E}(\ketbra{\bar{\psi}_{\a}})&=\sum_i E_i \ketbra{\bar{\psi}_{\a}} E^{\dagger}_i \\
&=\sum_i (G_i \otimes I_B) \ketbra{\bar{\psi}_{\a}}(G^\dagger_i \otimes I_B)+\ket{\phi^{\bot}_{{\a},i}}\!\bra{\bar{\psi}_\a}(G^\dagger_i \otimes I_B)+\mathrm{h.c.}+\ketbra{\phi^{\bot}_{{\a},i}}\ .
\label{Eq:code-detection}
\end{align}
\ees
Since $\ket{\bar{\psi}_{\a}} \in C$ and $C = A \otimes B$ (where $A$ is the gauge subsystem and $B$ is the  information subsystem), we have $\ket{\bar{\psi}_{\a}} = \ket{\bar{\psi}_{A,\a}} \otimes \ket{\bar{\psi}_{B,\a}}$, where $\ket{\bar{\psi}_{A,\a}} \in A$ and $\ket{\bar{\psi}_{B,\a}} \in B$. Thus the first term in Eq.~\eqref{Eq:code-detection} becomes $\sum_i (G_i \otimes I_B) \ketbra{\bar{\psi}_{\a}}(G^\dagger_i \otimes I_B) = \rho_{A,\a} \otimes \ketbra{\bar{\psi}_{B,\a}}$, where $\rho_{A,\a} = \sum_i G_i  \ketbra{\bar{\psi}_{A,\a}}G^\dagger_i$.

Now consider the observable $M \equiv P_C - P_{C^{\perp}}$; it has eigenvalue $+1$ ($-1$) for states in (orthogonal to) the codespace. Thus measuring $M$ is equivalent to detecting whether the measured state is in $C$ or in $C^{\perp}$. 
Clearly, measuring $M$ annihilates the off-diagonal term $\ket{\phi^{\bot}_{{\a},i}}\!\bra{\bar{\psi}_\a}(G^\dagger_i \otimes I_B)$ and its Hermitian conjugate. The post-measurement states are
\begin{equation}
\mathcal{E}(\ket{\bar{\psi}_{\a}}\bra{\bar{\psi}_{\a}})\stackrel{M}{\longmapsto}\left\{
\begin{array}{lll}
\frac{1}{p_+}{P_C\mathcal{E}(\ket{\bar{\psi}_{\a}}\bra{\bar{\psi}_{\a}})P_C} = \frac{1}{p_+}\rho_{A,\a}\otimes \ketbra{\bar{\psi}_{B,\a}} & \text{with prob. $p_+ = \Tr(\rho_{A,\a}) \braket{\bar{\psi}_{B,\a}|\bar{\psi}_{B,\a}}$}\\
&\\
\frac{1}{p_-}{P_{\mathcal{C}^{\bot}}\mathcal{E}(\ket{\bar{\psi}_{\a}}\bra{\bar{\psi}_{\a}})P_{\mathcal{C}^{\bot}}} = \frac{1}{p_-} \ketbra{\phi^{\bot}_{{\a},i}}& \text{with prob. $p_- = \braket{\phi^{\bot}_{{\a},i}|\phi^{\bot}_{{\a},i}}$}\\
\end{array}\right..
\end{equation}
Thus, if after measuring $M$ we obtain the outcome $+1$ corresponding to the projector $P_C$, the state is projected to the original information subsystem state $\ketbra{\bar{\psi}_{B,\a}}$ up to an irrelevant  transformation on the gauge subsystem. In this case no error took place. On the other hand, if the outcome $-1$ corresponding to the projector $P_{\mathcal{C}^{\perp}}$ is obtained, then we know that an error has happened. This shows that Eq.~(6) is sufficient for error detection using subsystem codes.

\section{A semi-distance and its relation to distinguishability in logical subsystems}
\label{app:semidist}

In the main text we argued that in the infinite penalty case measurement outcomes do not change despite the fact that the state evolves under $U_{\mrp}(T) U_0(T)$ rather than $U_0(T)$. To see this, consider replacing a generalized measurement $\mathcal{M}$ with measurement operators $\{M_m\}$ after evolving the initial state $\rho(0)$ subject to $H_S(t)$, by the encoded version $\overline{\mathcal{M}}=\{\overline{M}_m\}$ after evolving the encoded initial state $\overline{\rho}(0)$ using the encoded Hamiltonian $\overline{H}_S(t)$. Using Theorem 1, $U_V(T)\overline{\rho}(0)U^\dagger_V(T)\stackrel{E_{\mrp}\to\infty}{\longrightarrow} U_{\mrp}^\dagger(T) U_0^\dagger(T)\overline{\rho}(0)U_0(T) U_{\mrp}(T) \equiv \overline{\rho}(T)$, so the probability of outcome $m$ is 
$\Tr[\overline{\rho}(T) \overline{M}^\dagger_m \overline{M}_m  ] 
=\Tr[U_0^\dagger(T)\overline{\rho}(0)U_0(T) \overline{M}^\dagger_m \overline{M}_m ]$,
where we used $[U_{\mrp},\overline{M}_m]=0$ since the gauge operators commute with the logical operators. Thus, measurement outcomes do not change despite the fact that the state evolves under $U_{\mrp}(T) U_0(T)$ rather than $U_0(T)$.

To make this more precise and to relate it to the bounds derived for the finite penalty case, one can define a semi-distance that quantifies state distinguishability using measurements restricted to the logical subsystem.  

First we limit our discussion to the stabilizer subsystem setting; later we show how this can be generalized to general subsystem codes.

Let us denote the unitary that implements the encoding in the stabilizer formalism by $U_{\mathrm{enc}}$. This unitary maps an initial state, consisting of $s$ ancillas in the $\ket{0}$ state, some arbitrary state on $r$ gauge qubits $\ket{\phi}$, and a $k$-qubit information-carrying state $\ket{\psi}$, to an encoded state over $n=s+r+k$ physical qubits: $\ket{\overline{\psi}}=U_{\mathrm{enc}} (\ket{0}^{\otimes s} \ket{\phi} \ket{\psi})$. This unitary also converts the single Pauli operators on $n$ qubits  to the corresponding $s$ generators of the stabilizer, $r$ gauge generators, and $2k$ logical operators of the code \cite{shin2012codeword}.

We are interested in bounding the distance between the following two states:
\ba
\rho_{\mathrm{full}} = U_V(T) \overline{\rho} U^\dagger_V(T)\ ,\quad \rho_{\mathrm{ideal}}= U_0(T) \overline{\rho} U^\dagger_0(T)\ ,
\ea
where $\overline{\rho} = \ketbra{{\overline{\psi}}}$ is an initial state prepared in the support of $P$. The state $\rho_{\mathrm{full}}$ represents the evolution of this initial state under the effect of the total system-bath Hamiltonian, while $\rho_{\mathrm{ideal}}$ is the state resulting from the ideal, fully decoupled evolution subject purely to $H_0$. However, the gauge degrees of freedom need to be removed before a meaningful distance can be computed, since the state of the gauge subsystem is completely arbitrary. To account for this we need to define an appropriate distance measure, namely:
\ba
d(\rho,\sigma)=\frac{1}{2}\norm{\Tr_{\mathrm{gauge}} U_{\mathrm{enc}}^{\dagger} \rho  U_{\mathrm{enc}}-\Tr_{\mathrm{gauge}} U_{\mathrm{enc}}^{\dagger}  \sigma  U_{\mathrm{enc}}}_1 \ ,
\label{26}
\ea
which is the trace distance between states after tracing out the $r$ gauge qubits, and so it quantifies state distinguishability after a measurement of the logical subsystem. 
The unencoding transformation $U_{\mathrm{enc}}^{\dagger}$ is inserted in this definition in order to ensure a tensor-product structure between the gauge qubits and the rest.

We are thus interested in bounding $d(\rho_{\mathrm{full}},\rho_{\mathrm{ideal}})$, and proceed to do so. For our purposes it suffices to consider unitary operators that only act on the gauge degrees of freedom. Thus, we define a unitary family 
\ba
U_G = \{u_g :  U_{\mathrm{enc}}^{\dagger}u_g U_{\mathrm{enc}}=u_{g'}^{\mathrm{gauge}}\}\ ,
\ea
i.e., all the unitary operators $u_g$ whose effect on the codespace after unencoding is a unitary that has support only on the gauge qubits.

For elements of $U_G$ we have
\ba
\Tr_{\mathrm{gauge}} U_{\mathrm{enc}}^{\dagger} u_g \sigma u_g^{\dagger} U_{\mathrm{enc}}&=&\Tr_{\mathrm{gauge}} U_{\mathrm{enc}}^{\dagger} u_g U_{\mathrm{enc}} U_{\mathrm{enc}}^{\dagger} \sigma   U_{\mathrm{enc}}^{\dagger} U_{\mathrm{enc}}  u_g^{\dagger} U_{\mathrm{enc}} \notag \\
&=&\Tr_{\mathrm{gauge}} u_{g'}^{\mathrm{gauge}} U_{\mathrm{enc}}^{\dagger} \sigma  U_{\mathrm{enc}}^{\dagger} u_{g'}^{\mathrm{gauge}\dagger }\notag \\
&=&\Tr_{\mathrm{gauge}} U_{\mathrm{enc}}^{\dagger} \sigma   U_{\mathrm{enc}}\ ,
\ea
and so  
\ba
\forall u_g \in U_G : \quad d(\rho,\sigma)=d(\rho,u_g \sigma u_g^{\dagger}) \ .
\ea
From this we conclude that:
\ba
d(\rho,\sigma) &=& \frac{1}{2} \min_{u_g \in U_G} \norm{\Tr_{\mathrm{gauge}} U_{\mathrm{enc}}^{\dagger}  \rho  U_{\mathrm{enc}}-\Tr_{\mathrm{gauge}} U_{\mathrm{enc}}^{\dagger} u_g   \sigma  u_g^{\dagger} U_{\mathrm{enc}}}_1\\
&\leq& \frac{1}{2} \min_{u_g \in U_G} \norm{ U_{\mathrm{enc}}^{\dagger}  \rho  U_{\mathrm{enc}}- U_{\mathrm{enc}}^{\dagger} u_g   \sigma  u_g^{\dagger} U_{\mathrm{enc}}}_1 \nonumber \\
&=& \frac{1}{2}\min_{u_g \in U_G} \norm{  \rho  - u_g  \sigma  u_g^{\dagger} }_1 \ .
\label{31}
\ea
Now note that $U_{\mrp}(T)\in U_G$, as it is a unitary generated by a linear combination of elements of the gauge group. Therefore we have
\bes
\label{32}
\ba
d(\rho_{\mathrm{full}},\rho_{\mathrm{ideal}}) &\leq& \frac{1}{2}\min_{u_g \in U_G} \norm{  \rho_{\mathrm{full}}  - u_g  \rho_{\mathrm{ideal}}  u_g^{\dagger} }_1 \\
&\leq&\frac{1}{2}\min_{u_g \in U_G} \norm{ U_V \overline{\rho} U_V^{\dagger}  - u_g U_0 \overline{\rho}U_0^{\dagger} u_g^{\dagger} }_1 \\
\label{32c}
&\leq& \frac{1}{2}\norm{ U_V \overline{\rho} U_V^{\dagger} - U_0 U_{\mrp} \overline{\rho} U_0^{\dagger} U_{\mrp}^{\dagger} }_1,
\ea
\ees

Theorem~1 involves the operator norm $\norm{ U_V P - U_0U_{\mrp} P}$, so another step is required in order to connect the bound on $d(\rho_{\mathrm{full}},\rho_{\mathrm{ideal}})$ with Theorem~1.
This can be easily done using
\ba
\frac{1}{2}\norm{U_V \rho U_V^\dagger - U_W \rho U_W^\dagger}_1 \leq \norm{U_V P- U_W P}\ ,
\ea
where it is assumed that the initial state $\rho$ is in the support of $P$ (for completeness the bound is derived in the next section). Thus, we have obtained the desired bound as:
{
\ba
d(\rho_{\mathrm{full}},\rho_{\mathrm{ideal}}) \leq \norm{ U_V P - e^{-ic}U_0U_{\mrp} P}\ .
\ea
}
Of course this distance also goes to zero in the large penalty limit.

\subsubsection{General subsystem case}
To generalize the distance we have defined to general subsystem codes we note that, unlike in the case of stabilizer codes, $U_{\mathrm{enc}}^{\dagger}$ only produces the tensor structure in the codespace ($\mathcal{H}=C \oplus C^\perp$, where $C = {A} \otimes {B}$.) In this case we can modify the defined distance to trace out the gauge degrees of freedom only in the codespace:
\ba \label{def:dgeneral}
d(\rho,\sigma)=\frac{1}{2}\norm{\Tr_{\mathrm{gauge}} U_{\mathrm{enc}}^{\dagger} P_C \rho P_C U_{\mathrm{enc}}-\Tr_{\mathrm{gauge}} U_{\mathrm{enc}}^{\dagger} P_C  \sigma P_C  U_{\mathrm{enc}}}_1 \ .
\ea
As in Theorem~2,  here we also assume that $[P,H_{\mrp}]=0$.
All the steps described for the stabilizer case above can be repeated with small modifications. We define the unitary family 
\ba
U_G = \{u_g : [u_g,P_C]=0\, , \,\, U_{\mathrm{enc}}^{\dagger}u_g U_{\mathrm{enc}}=u_{g'}^{\mathrm{gauge}}\}\ ,
\ea
i.e., all the unitary operators $u_g$ that both commute with $P_C$ and whose effect on the codespace after unencoding is a unitary that has support only on the gauge subsystem. Using this one can easily show that
\ba
\forall u_g \in U_G : \quad d(\rho,\sigma)=d(\rho,u_g \sigma u_g^{\dagger}) \ .
\ea
From this we conclude that:
\ba
d(\rho,\sigma) 
&\leq& \frac{1}{2}\min_{u_g \in U_G} \norm{ P_C \rho P_C - u_g P_C \sigma P_C u_g^{\dagger} }_1 \ .
\label{31'}
\ea
Thus, we have:
\bes
\label{32'}
\ba
d(\rho_{\mathrm{full}},\rho_{\mathrm{ideal}}) &=& \frac{1}{2}\norm{\Tr_{\mathrm{gauge}} U_{\mathrm{enc}}^{\dagger} P_C U_V \overline{\rho} U^\dagger_V P_C U_{\mathrm{enc}}-\Tr_{\mathrm{gauge}} U_{\mathrm{enc}}^{\dagger} P_C  U_0 \overline{\rho} U^\dagger_0 P_C  U_{\mathrm{enc}}}_1 \\
\label{32b}
&\leq&\frac{1}{2}\min_{u_g \in U_G} \norm{ P_C U_V \overline{\rho} U_V^{\dagger} P_C - u_g P_C U_0 \overline{\rho}U_0^{\dagger}P_C u_g^{\dagger} }_1 \\
\label{32c'}
&\leq& \frac{1}{2}\norm{ P_C U_V \overline{\rho} U_V^{\dagger} P_C - P_C U_0 U_{\mrp} \overline{\rho} U_0^{\dagger} U_{\mrp}^{\dagger} P_C }_1\\
&\leq&\frac{1}{2}\norm{ U_V \overline{\rho} U_V^{\dagger} - U_0 U_{\mrp} \overline{\rho} U_0^{\dagger} U_{\mrp}^{\dagger}}_1 \\
&\leq& \norm{ U_V P - e^{-ic}U_0U_{\mrp} P}\ ,\ea
\ees
where in the last step we assumed that the initial state $\overline{\rho}$ is in the support of $P$.

We remark that 
as we are just comparing the part of the states in the codespace, the distance $d$ defined in Eq.~\eqref{def:dgeneral} can vanish for two different states that are \emph{both} outside the codespace. However, when one of the states $\rho_C$  is in the codespace, while the other state $\sigma$ is not necessarily in the codespace, having $d(\rho_C,\sigma) \leq \epsilon$ guarantees that $\Tr(P_C \sigma) \geq 1- \epsilon$. Thus, if the distance $d$ between a general state and a state in the codespace is small, it is guaranteed that $\sigma$ is close to the codespace.

We can relate this to the problem of optimally distinguishing states, with general prior probabilities: For two general states with nonzero support in the codespace, we can rewrite the distance definition as
\ba
d(\rho,\sigma)=\frac{1}{2}\norm{p \rho' -q \sigma' }_1 \ ,
\ea
where we have defined $p=\Tr P_C \rho$ and $\rho' =\Tr_{\mathrm{gauge}}  U_{\mathrm{enc}}^{\dagger} P_C \rho P_C U_{\mathrm{enc}} / p$, and likewise, $q=\Tr P_C \sigma$ and $\sigma' =\Tr_{\mathrm{gauge}}  U_{\mathrm{enc}}^{\dagger} P_C \sigma P_C U_{\mathrm{enc}} / q$. This quantity has an operational meaning connected to the minimum error of distinguishing states $\rho'$ and $\sigma'$ with prior probabilities $p$ and $q$: $p_{\mathrm{error}}= \frac{1}{2}\left(1-\norm{p \rho' - q \sigma'}_1\right)$ \cite{Wilde:book}.

\subsection{Trace distance of states and operator norm of evolutions}
For pure states, there exists a tight relation between fidelity and trace distance of states \cite{nielsen2010quantum}:
\ba
D(\ketbra{\psi_1} , \ketbra{\psi_2})=\sqrt{1-|\braket{\psi_1|\psi_2}|^2} \leq \norm{\ket{\psi_1}-\ket{\psi_2}}\ .
\ea
%
%
Thus, if $\ket{\psi_1}= U_V  \ket{\psi_0}$ and $\ket{\psi_2}= U_W  \ket{\psi_0}$ for some pure state $\ket{\psi_0}$, and $P\ket{\psi_0}=\ket{\psi_0}$
we have: 
\ba 
\label{eq:Tracenorm}
D(\ketbra{\psi_1} , \ketbra{\psi_2}) \leq {\norm{(U_V -U_W )\ket{\psi_0}}}={\norm{(U_V -U_W )P \ket{\psi_0}}} \leq {\norm{U_V P-U_W P }}\ ,
\ea
where we used the definition of the operator norm in the last inequality.

The same method extends to the case in which the initial state is a mixed state, by decomposing the initial state into an ensemble of pure states: $\rho = \sum_i p_i \ketbra{\psi_i}$, where $\{p_i\}$ is a probability distribution. If this state is in the support of $P$, then:
\bes
\label{36}
\begin{align}
D(U_V \rho U_V^\dagger , U_W \rho U_W^\dagger) &= \frac{1}{2}\norm{\sum_i p_i  (U_V \ketbra{\psi_i} U_V^\dagger - U_W \ketbra{\psi_i} U_W^\dagger)}_1 \\
& \leq \frac{1}{2}\sum_i p_i \norm{  (U_V \ketbra{\psi_i} U_V^\dagger - U_W \ketbra{\psi_i} U_W^\dagger)}_1\\
&= \sum_i p_i D(U_V \ketbra{\psi_i} U_V^\dagger , U_W \ketbra{\psi_i} U_W^\dagger)\\
&\leq \sup_{\ket{\psi_i}} D(U_V \ketbra{\psi_i} U_V^\dagger , U_W \ketbra{\psi_i} U_W^\dagger)\\
\label{36e}
&\leq \norm{U_V P- U_W P}\ ,
\end{align}
\ees
where we used Eq.~\eqref{eq:Tracenorm} in Eq.~\eqref{36e}.

\section{Generalized Bacon-Shor codes using Bravyi's $A$-matrix construction}

Here we briefly review the construction and properties of the generalized Bacon-Shor codes introduced in Ref.~\cite{PhysRevA.83.012320}.
Let $A$ be a square binary matrix. Following Bravyi, we associate a subsystem code to this matrix. Each non-zero element of $A$ represents a physical qubit, so the number of physical qubits, $n$, is just the Hamming weight $|A|$. The gauge group of the code, $\mathcal{G}$, is generated by:
\begin{enumerate}
\item Gauge generators $X_a X_b$, for each pair of qubits $a,b$ in the same row,
\item Gauge generators $Z_a Z_b$, for each pair of qubits $a,b$ in the same column.
\end{enumerate}
Note that, by definition, the generators can be overcomplete. It suffices to retain generators for consecutive pairs.

Given the gauge group, in principle all properties of the code can be derived using: 
\bes
\begin{align}
\mathcal{S}&=\mathcal{G} \cap C(\mathcal{G})\quad (\textrm{stabilizer group})\ ,\\
\mathcal{L}&= C(\mathcal{S})/\mathcal{G} \quad (\textrm{dressed logical operators})\ ,\\
d&=\min_{P \in C(\mathcal{S})/\mathcal{G} }{|P|} \quad (\textrm{code distance})\ .
\end{align}
\ees
But for this specific construction, all these properties can be directly related to properties of the $A$ matrix over the binary field ${\mathbb{F}}_2$. As shown in Ref.~\cite{PhysRevA.83.012320} (Theorem 2 there) the properties of the code are: 
\ba
[[n,k,d]]=[[|A|, \mathrm{rank}(A), \min\{d_{\mathrm{row}}, d_{\mathrm{col}}\}]]
\ea
The rank of $A$ over ${\mathbb{F}}_2$ is equal to half the number of logical operators of the code, $k$. The code distance is the minimum weight of the non-zero vectors in the row space and the non-zero vectors in the column space.

We present the $A$ matrices and corresponding gauge groups for the examples given in the main text.
We use the notation $X_{i,j}$ to represent the $X$ operator acting on the qubit located in the $i^\mathrm{th}$ row and $j^\mathrm{th}$ column (in the main text we used one index to label the qubits for simplicity).

\begin{enumerate}[label=(\roman*)]
\item $[[4,1,2]]$ code: $A=\begin{bmatrix}
  1 & 1\\
  1 & 1\\
  \end{bmatrix}
$, so
\ba
\mathcal{G}=\braket{X_{1,1}X_{1,2}, X_{2,1}X_{2,2},Z_{1,1}Z_{2,1},Z_{1,2}Z_{2,2}}\ .
\ea
This is the code for encoding one qubit. If we use this code to encode all $N$ qubits of the system (block encoding), the corresponding $A$ becomes block diagonal with $\begin{bmatrix}
  1 & 1\\
  1 & 1\\
\end{bmatrix}$ on the diagonal.

\item $[[8,3,2]]$ code for the adiabatic swap gate:
$A=\begin{bmatrix}
  1 & 1 & 0 & 0\\
  0 & 1 &1 & 0\\
  0 & 0 & 1 & 1 \\
  1 & 0 & 0 & 1
\end{bmatrix}
$, so:
\ba
\mathcal{G}=\braket{X_{1,1}X_{1,2}, X_{2,2}X_{2,3},X_{3,3}X_{3,4},X_{4,1}X_{4,4},Z_{1,1}Z_{4,1},Z_{1,2}Z_{2,2},Z_{2,3}Z_{3,3},Z_{3,4}Z_{4,4}}\ .
\ea

\item Linear chain:
\ba
A=
\begin{bmatrix}
  1 &1 &0 & 0 \dots &0 &0 \\
  0 &1 &1 & 0\dots&0 &0\\
  0 &0 &1 & 1\dots&0 &0\\
 \vdots & \vdots\\
   0 &0 &0 & 0...&1 &1\\
  1 &0 &0 & 0...&0 &1
\end{bmatrix}_{a \times a} \ .
\ea
This corresponds to a $[[|A|, \mathrm{rank}(A), \min(d_{\mathrm{row}},d_{\mathrm{col}})]]=[[2a,a-1,2]]$ code, $a=N+1$, and the generators of $\mathcal{G}$ are:
\bes
\label{eq:57}
\begin{align}
1 \leq \forall i \leq a-1: \quad  &X_{i,i} X_{i,i+1}\\
& X_{a,1}X_{a,a}\\
1 \leq \forall  i \leq a-1: \quad  &Z_{i,i+1} Z_{i+1,i+1} \\
&Z_{1,1} Z_{a,1} \ .
\end{align}
\ees
The logical operators can be chosen to be 
\bes
\begin{align}
1 \leq \forall i \leq a-1: \quad & \overline{X}_i= X_{i,i+1} X_{i+1,i+1}\\
1 \leq \forall i \leq a-1: \quad & \overline{Z}_i= \prod_{j=1}^{i} Z_{j,j} Z_{j,j+1}\ .
\end{align}
\ees
Thus $2 \leq \forall i \leq a-1 : \overline{Z}_{i-1} \overline{Z}_{i} =Z_{i,i}Z_{i,i+1}$, a $2$-local physical interaction. 

The result in Eq.~(14) follows: $
\overline{H}_S(s)
=(1-s) \sum_{i=1}^N{\overline{X}_i}+ s \sum_{i=1}^{N-1}J_i{\overline{Z}_i \overline{Z}_{i+1}}
=(1-s) \sum_{i=1}^{N} {X_{2i} X_{2i+1}}+ s \sum_{i=1}^{N-1} J_i {Z_{2i+1} Z_{2i+2}}$.

\end{enumerate}

\section{Checking Condition~1 for the $[[4,1,2]]$ code}
\label{app:check412}

Recall that 
\bes
\ba
\mathcal{S}&=&\braket{X_1 X_2 X_3 X_4 ,  Z_1 Z_2 Z_3 Z_4}  \\
\mathcal{G}&=&\braket{X_3 X_4,Z_2 Z_4,X_1 X_2, Z_1 Z_3} \equiv \braket{\{g_i\}_{i=1}^4}\\
h_{\mrp}&=&E_{\mrp} \sum_{i=1}^4  g_i\ .
\ea
\ees
First note that $h_{\mrp} P_C=2 E_{\mrp} (g_1 +g_2) P_C$. The eigenvalues of this matrix are $\pm 2E_{\mrp} \sqrt{2}$ (note that the spectrum is equivalent to the spectrum of $XIII+ZIII$, as there exists a unitary transformation between the terms).

Next, observe that a single Pauli $X$ at any of the four locations commutes with both $g_1,g_3$ and it commutes with one of $g_2,g_4$ and anticommutes with the other one. So, for $\sigma_j$ being $X$ at any location we have $\sigma_j h_{\mrp} \sigma_j P_C=2 E_{\mrp} g_1 P_C$. Similarly, for $\sigma_j$ being $Z$ at any location we have $\sigma_j h_{\mrp} \sigma_j P_C=2 E_{\mrp} g_2 P_C$. However, if $\sigma_j=Y$ then at any location it commutes with one of $g_1,g_3$ and anti-commutes with the other one, and also commutes with one of $g_2,g_4$ and anti-commutes with the other one; thus $\sigma_j h_{\mrp} \sigma_j P_C=0$. 
Therefore the eigenvalues are either $\pm 2E_{\mrp}$ or $0$.

\section{Details for the protected Ising chain in a transverse field}
\begin{figure}
\begin{center}
\includegraphics[height=0.25\textheight]{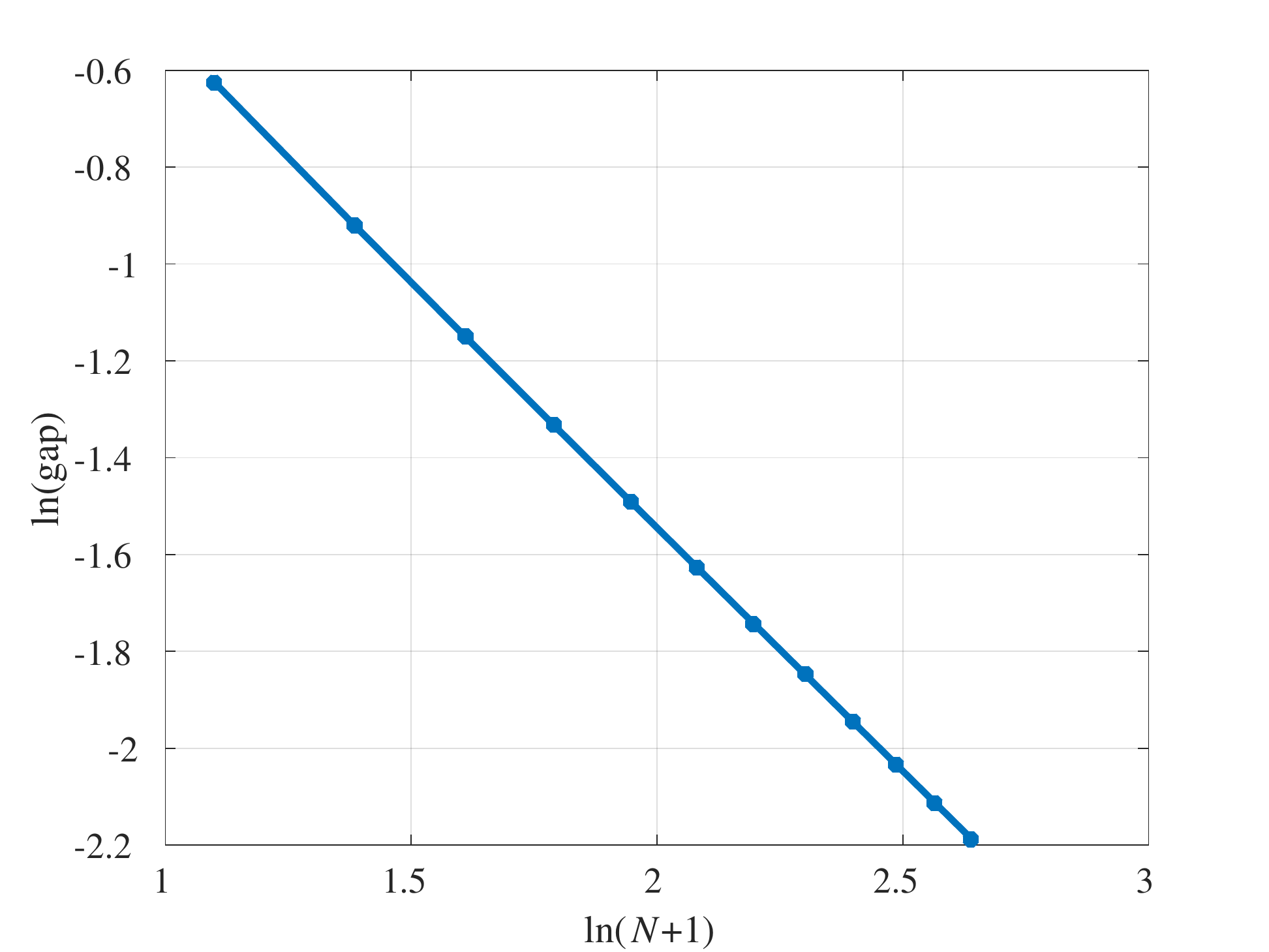}
\caption{Numerically computed gap of the penalty Hamiltonian for the protected Ising chain in a transverse field, as a function of the number of qubits, given the scaling relation $\mathrm{gap} \sim 1/(N+1)$.}
\label{Fig:1}
\end{center}
\end{figure}

The spectrum of the penalty Hamiltonian given in the main text, $H_{\mathrm p}= -(\sum_{i=1}^{N+1} {X_{2i-1} X_{2i}} + \sum_{i=1}^{N} {Z_{2i} Z_{2i+1}+Z_{2N+2}} Z_1)$ [acting on $2(N+1)$ qubits] can be found by considering the spectrum of the following Hamiltonian, with $s_x$ and $s_z$ set to $\pm 1$:
\ba
-(\sum_{i=1}^{N}{{X}_i} + s_x \prod_{i=1}^{N} {X}_i + {Z}_1 + \sum_{i=1}^{N-1}{{Z}_i {Z}_{i+1}} +s_z {Z}_N)\ .
\label{eq:H-transformed}
\ea
The minimum energy of this Hamiltonian appears in the $s_x=s_z=+1$ sector and so the ground subspace is in the codespace, as required. Figure~\ref{Fig:1} shows how the gap of the penalty Hamiltonian changes as a function of the number of qubits. This gap is proportional to $(N+1)^{-1}$.

To arrive at Eq.~\eqref{eq:H-transformed} we define
\bes
\begin{align}
1 \leq \forall i \leq N: \, \hat{X}_i&=X_{2i-1} X_{2i}\\
\hat{Z}_1&=Z_{2N+2}Z_1 \\
2 \leq \forall i \leq N :  \,  \hat{Z}_i&= Z_{2N+2}Z_{1} \Pi_{j=1}^{i-1}{Z_{2j}Z_{2j+1}}\ .
\end{align}
\ees
Note that these operators are similar in form but different in indexing to the logical operators in Eq.~\eqref{eq:57}. With the stabilizers $S_x$ ($S_z$) being the product of all $X$ ($Z$) Pauli operators on all $2(N+1)$ qubits, the original penalty Hamiltonian can be rewritten as
\ba
-(\sum_{i=1}^{N}{{\hat{X}}_i} + S_x \prod_{i=1}^{N} {\hat{X}}_i + {\hat{Z}}_1 + \sum_{i=1}^{N-1}{{\hat{Z}}_i {\hat{Z}}_{i+1}} +S_Z {\hat{Z}}_N)
\ea
Now, as these two stabilizers commute with all the terms, we can diagonalize the Hamiltonian in the corresponding $\pm 1$ sectors separately.
As these hatted operators satisfy the same algebra as single Pauli operators the spectrum will be the same. 


\end{document}